\documentclass[12pt,twoside]{article}
\usepackage{fleqn,espcrc1}

\title{\bf On the Mixing Amplitude of $J/\psi$ and Vector Glueball $O$}

\author{Chuan-Tsung Chan\thanks{E-mail address: ctchan@phys.ntu.edu.tw}
        and 
        Wei-Shu Hou\thanks{E-mail address: wshou@phys.ntu.edu.tw} \\
        \vspace{0.4cm}
        Department of Physics, National Taiwan University,\\      
        Taipei, 10617, Taiwan, ROC}
       
\begin{document}

\maketitle

\begin{abstract}
  We study the mixing angle $\theta_{O\psi}$ and mixing amplitude
$f_{O\psi}$ of $J/\psi$ and vector glueball $O$, in the framework of 
potential models of heavy quarks and constituent gluons.
  While the state vectors of $J/\psi$ and $O$ are constructed from
the wave functions of few-body Schroedinger equations, the mixing
dynamics is governed by perturbative QCD.
  We obtain a value of the mixing angle of $|\tan \theta_{O\psi}| \approx 
0.015$ and the mixing amplitude of $|f_{O\psi}(m_{O\psi}^2)| \approx 0.008 
\mbox{ GeV}^2$, which is compatible with phenomenological analysis. 
\end{abstract}

\section{Introduction}
 
  Among various solutions attempting to resolve the ``$\rho \pi$ anomaly" in 
charmonium decays \cite{Olsen:1997qi} \cite{Tuan:1999ig}, the resonance 
enhancement model, as proposed by Hou and Soni \cite{Hou:1983kh} and later 
generalized by Brodsky, Lepage, and Tuan \cite{Brodsky:1987bb}, requires
the existence of a vector glueball $O$ \cite{Freund:1975pn}. 
  Since both the rest energy $m_O - i \ {\Gamma_O / 2}$ and the mixing 
amplitudes $f_{O\psi}, f_{O \psi'}$ with charmonium $J/\psi$ and $\psi'$
are needed in the analysis of ``$\rho \pi$ anomaly "\cite{Hou:1997qk}, it is 
necessary to study these issues within a hadronic model and henceforth provide
some quantitative information.
  A naive approach \cite{Lo} based on a constituent model of gluons 
\cite{Cornwall:1982zr} was used earlier to study the glueball spectrum in the
pure Yang-Mills gauge theory.
  In the present study \cite{ctws}, we extend this framework by including 
the quark-gluon interaction and calculate the mixing angle $\theta_{O\psi}$
within a nonrelativistic approximation.

\section{Setup of the Problem}

  In the non-relativistic framework, we can calculate the mixing angle 
$\tan \theta_{O\psi} (\vec Q)$ between the physical composites $J/\psi$,
$O$, and un-mixed hadrons $c \bar c$ and $ggg$ via the evolution operator
$U(T, -T) = e^{ - 2 i H T}$,
\begin{equation}
\left( \begin{array}{c} | J/\psi (\vec Q) \rangle \\
                        |   O    (\vec Q) \rangle
       \end{array} \right)_{NR} =
\left( \begin{array}{rr}  \cos\theta_{O\psi}(\vec Q)\hspace{0.5cm}
                          \sin\theta_{O\psi}(\vec Q) \\  
                        - \sin\theta_{O\psi}(\vec Q)\hspace{0.5cm}
                          \cos\theta_{O\psi}(\vec Q)
       \end{array} \right) \
\left( \begin{array}{c} | c \bar c (\vec Q) \rangle \\
                        |   ggg    (\vec Q) \rangle
       \end{array} \right)_{NR} \ ,
\end{equation}
\begin{equation}
- \tan \theta_{O\psi}(\vec Q)
= \lim_{T \rightarrow \infty \cdot e^{i \epsilon}}
  \frac{\langle      ggg (\vec Q) | U(T,-T) | c \bar c (\vec Q) 
        \rangle_{NR}} 
  { \ \ \langle c \bar c (\vec Q) | U(T,-T) | c \bar c (\vec Q) 
        \rangle_{NR}} 
\approx \langle ggg (\vec Q) | U(T,-T) | c \bar c (\vec Q) \rangle_{NR} .
\label{eq:mixangle}
\end{equation}
  Neglecting the decay widths of both $J/\psi$ and $O$, the mixing angle 
$\theta_{O\psi}$ is then related to a relativistically normalized mixing 
amplitude $f_{O\psi}$, by
\begin{eqnarray}
f_{O\psi}(q^2) &=& \left[ (\gamma \delta - \alpha \beta) q^2
                    + (\alpha \beta  m_{ggg}^2 -
                       \gamma \delta m_{c \bar c}^2)
                   \right]
                   \sin \theta_{O\psi} \cos \theta_{O\psi} \\
         &\approx& \left[ (\delta - \beta)  q^2
                   + ( \beta m_{ggg}^2 - \delta m_{c \bar c}^2 )
                   \right] \theta_{O\psi} \ ,
\end{eqnarray}
where in the second line we assume a small mixing angle and keep terms only 
to first order in $\theta_{O\psi}$, and the state-dependent normalization 
factors are
\begin{equation}
\begin{array}{lclclclclclc}
\alpha(\vec Q) &\equiv& \sqrt{\frac{\omega_{J/\psi}(\vec Q) }
                                   {\omega_{c \bar c}(\vec Q) }}      
       & = &    \left( \frac{{\vec Q}^2 + m_{J/\psi}^2}
                            {{\vec Q}^2 + m_{c \bar c}^2} \right)^{1/4} ,
& &
\beta(\vec Q)  &\equiv& \sqrt{\frac{\omega_{J/\psi}(\vec Q) }
                                   {\omega_{ggg}(\vec Q)}}
       & = & \left( \frac{{\vec Q}^2 + m_{J/\psi}^2}
                         {{\vec Q}^2 + m_{ggg}^2} \right)^{1/4} ,\\
\gamma(\vec Q) &\equiv& \sqrt{\frac{\omega_O(\vec Q)}
                                   {\omega_{ggg}(\vec Q) }}
       & = & \left( \frac{{\vec Q}^2 + m_O^2}
                         {{\vec Q}^2 + m_{ggg}^2} \right)^{1/4} ,
& &   
\delta(\vec Q) &\equiv& \sqrt{\frac{\omega_O(\vec Q)}
                                   {\omega_{c \bar c}(\vec Q) }}
       & = & \left( \frac{{\vec Q}^2 + m_O^2}
                         {{\vec Q}^2 + m_{c \bar c}^2} \right)^{1/4} . \\
\end{array}
\end{equation}

  With these definitions, our task is to calculate the transition amplitude 
Eq.(\ref{eq:mixangle}) in the framework of constituent models of charm 
quarks and gluons. 
  Specifically, we need to construct the state vectors of $c \bar c$ and 
$ggg$ and then calculate the annihilation amplitude between these 
constituents. 
  The transition amplitude between $c \bar c$ and $ggg$ composites can 
then be expressed as a convolution of the bound state wave functions and 
the annihilation amplitude among quark/gluon constituents.

\section{Details of the calculations}

  In this section, we outline the basic ingredients in our calculations 
of the transition amplitude. 
  Our normalizations and conventions follow that of \cite{Nachtmann:1990ta}.
\begin{enumerate}
 \item The state vector of the $c \bar c$ particle:
\begin{eqnarray}
   | c \bar c (\vec P, \vec \lambda) \rangle_{NR} &=&
       \frac{1}{\sqrt V} \
  \int \frac{d^3 {\vec p}_1}{(2 \pi)^3 \sqrt{2 \omega_c({\vec p}_1)}}
  \int \frac{d^3 {\vec p}_2}{(2 \pi)^3 \sqrt{2 \omega_c({\vec p}_2)}} \
    (2 \pi)^3     \delta^{(3)}( \vec P - {\vec p}_1 - {\vec p}_2 ) 
       \nonumber  \\ &\times& \
      {\tilde \Psi}_{c \bar c} (\frac{{\vec p}_1 - {\vec p}_2}{2}) \
    S_{rs}( \vec \lambda ) \ U^{\alpha \beta}    
  \ b^{\alpha \dagger}_r ({\vec p}_1)
  \ d^{\beta  \dagger}_s ({\vec p}_2) | 0 \rangle ,
\end{eqnarray}
where ${\tilde \Psi}_{c \bar c}(\vec p)$ is the momentum space wave 
function of $c \bar c$. 
  The spin and color wave function are given by 
$S_{rs}( \vec \lambda ) \equiv \frac{1}{\sqrt 2} \vec \lambda \cdot ( {\vec 
\sigma} \ \epsilon )_{rs}$, 
$U^{\alpha \beta} \equiv \frac{1}{\sqrt 3} \delta^{\alpha \beta}$,
respectively.

\item The state vector of the $ggg$ particle:
\begin{eqnarray}
 | g g g (\vec K, \vec \zeta) \rangle_{NR} &\sim& \
   \frac{1}{\sqrt V} \
   \left[ \ \Pi_{i=1}^3 \int
   \frac{d^3 {\vec k}_i}{(2 \pi)^3 \sqrt{2 \omega_g ({\vec k}_i)}} \ \right]
   \ (2 \pi)^3 \delta^{(3)}(\vec K - \sum_{i=1}^3 {\vec k}_i ) \
   \nonumber \\
  &\times& {\tilde \Phi}_{ggg} ({\vec k}_i; \vec K)
   \ T_{mpq}(\vec \zeta)
   \ V^{abc} \frac{1}{\sqrt 6} 
   \ G^{a \dagger}_m ({\vec k}_1)
   \ G^{b \dagger}_p ({\vec k}_2)
   \ G^{c \dagger}_q ({\vec k}_3) | 0 \rangle \ .
\end{eqnarray}
  
  Without solving the three body problem, we take a trial wave function 
of vector glueball $ggg$,
\begin{equation}
 {\tilde \Phi}_{ggg}({\vec k}_1,{\vec k}_2,{\vec k}_3; \vec K) \equiv
  \left( \frac{6 \pi}{a^2 m^2_g} \right)^{3/2} \! 
  \exp \left[ \ - \frac{1}{2 a^2 m^2_g} \left( - {\vec k}_1 {\vec k}_2
                                               - {\vec k}_2 {\vec k}_3
                                               - {\vec k}_3 {\vec k}_1
                           + \frac{{\vec K}^2}{3} \right) \ \right],
\end{equation}
and the variational parameter $a$ is fixed at $0.64$ to obtain a stable 
glueball spectrum \cite{Lo}. 
  The totally symmetric color wave function is given by $V^{abc} \equiv 
\sqrt{\frac{3}{40}} d^{abc}$.
  Please refer to \cite{ctws} for the definitions of creation operators of the
constituent gluons $G_m^{a \dagger} (\vec k)$ and the spin wave function 
$T_{mpq}(\vec \zeta)$.  

 \item The annihilation amplitude:

  Using the Feynman rules for charm quarks and constituent gluons,
we can write down the annihilation amplitude corresponding to the process
$c + \bar c \leftrightarrow ggg$,
\begin{equation}
 {\cal A}_{c \bar c \leftrightarrow ggg} =
    \bar{{\it v}}_n ({\vec p}_2, s)
         {\it u}_m  ({\vec p}_1, r)
    G_{\mu  i}({\vec k}_1)
    G_{\nu  j}({\vec k}_2)
    G_{\rho l}({\vec k}_3)
    S_{rs}(\vec \lambda)
    T_{ijl}(\vec \zeta)
    U^{\alpha \beta} 
    V^{abc}
 {\it A}^{a b c,\mu \nu \rho}_{\alpha \beta, n m} ,
\end{equation}
where
\begin{eqnarray}
{\it A}^{a b c,\mu \nu \rho}_{\alpha \beta, n m} &\equiv& (-i g_s)^3
 \left[
 \frac{\lambda^a}{2} \frac{\lambda^b}{2} \frac{\lambda^c}{2}
 \right]_{\alpha \beta}
 \left[
 \gamma^\mu
 \frac{i}{{\not p}_1 - {\not k}_1 - m_c}
 \gamma^\nu
 \frac{i}{{\not p}_1 - {\not k}_1 - {\not k}_2  - m_c}
 \gamma^\rho
 \right]_{n m}
 \nonumber \\
&+& \mbox{6 permutations ,}
\end{eqnarray}
and the external legs for constituent gluons are defined as
$G^{\mu i}({\vec k}_1) \equiv - g^{\mu i} + \frac{k_1^\mu k_1^i}{m_g^2}$.

  To simplify the algebra, we perform a momentum expansion,
\begin{eqnarray}
 {\cal A}_{c \bar c \leftrightarrow ggg}({\vec p}_i;     {\vec k}_j) &=&
 {\cal A}_{c \bar c \leftrightarrow ggg}({\vec p}_i = 0; {\vec k}_j   = 0)
+ \left( {\vec p}_i \frac{\partial}{\partial {\vec p}_i} +
         {\vec k}_j \frac{\partial}{\partial {\vec k}_j} \right) \  
 {\cal A}_{c \bar c \leftrightarrow ggg} 
 |_{{\vec p}_i = 0; {\vec k}_j = 0}
  \nonumber \\
&+& \mbox{higher order terms in } \frac{p_i}{m_c}, \frac{k_j}{m_g}    
\ \approx \ \frac{1.06 \ g_s^3}{{\sqrt 3} m_g (m_g - 2 m_c)} .
\end{eqnarray} 
 
\item Mixing angle as a convolution:
 
  A generalized Fermi's Golden Rules No.2 can be applied to the transition 
amplitude and the mixing angle formula Eq.(\ref{eq:mixangle}).
  With appropriate normalizations, we have
\begin{eqnarray}
  & & - \tan \theta_{O\psi}(\vec P) =
     \int \frac{d^3 {\vec p}_1}{(2\pi)^3 \sqrt{2 \omega_c ({\vec p}_1)}}
     \int \frac{d^3 {\vec k}_1}{(2\pi)^3 \sqrt{2 \omega_g ({\vec k}_1)}}
     \int \frac{d^3 {\vec k}_2}{(2\pi)^3 \sqrt{2 \omega_g ({\vec k}_2)}}
     \nonumber \\ & & \times
     (2 \pi) \ \delta( \sum_{i=1}^{2} \omega_c ({\vec p}_i) -
                       \sum_{j=1}^{3} \omega_g ({\vec k}_j)  ) \
     \frac{{\tilde \Psi}_{c \bar c} ({\vec p}_1)             \
           {\cal A}( {\vec p}_1; 
           {\vec k}_1, {\vec k}_2 ) \
           {\tilde \Phi}_{ggg}^*    ({\vec k}_1,{\vec k}_2 )}
          { \sqrt{ 6 }
            \sqrt{ 2 \omega_c(\vec P - {\vec p}_1) }
            \sqrt{ 2 \omega_g(\vec P - {\vec k}_1 - {\vec k}_2) } } . 
\label{eq:con}
\end{eqnarray}
  Since our model is based on a nonrelativistic approximation, we can perform 
a momentum expansion on a convoluted wave function of the vector glueball 
$ggg$, $\int d^3 k {\cal A} {\tilde \Phi}_{ggg}^*$.   
  The momentum integration of the wave function of the $c \bar c$ then
generates a factor of $c \bar c$ spatial wave function at origin 
$g_s^3 \Psi_{c \bar c}(0)$, which can be extracted from the total hadronic 
decay rate of $J/\psi$ \cite{Nachtmann:1990ta}.
  Using the mass inputs $m_c = 1.5 \mbox{ GeV}$ and $m_g = 0.7 \mbox{ GeV}$,
and the variational parameter $a=0.64$, our calculation give a value of 
mixing angle with $\tan \theta_{O\psi} (\vec Q = 0) \approx 0.015$. 

\item Systematics of our approximations:

    In this mixing angle calculations, we have made several approximations:  
\begin{enumerate}
\item Only nearest states, namely, $c \bar c$ and $ggg$, are considered.  
\item Mixing dynamics between quarkonium and glueball is treated perturbatively
      in the strong coupling constant $\alpha_s$.
\item Assuming a nonrelativistic picture, we expand the annihilation
      amplitude \\ ${\cal A}_{c \bar c \leftrightarrow ggg}$ in powers of 
      constituent momenta and keep leading term only.
\item A variational solution of the vector glueball wave function is used for 
      the convolution formula Eq.(\ref{eq:con}).
\end{enumerate}
  For a discussion of possible improvement, see our paper \cite{ctws} for 
details. 
\end{enumerate}
  
\section{Summary and conclusions}

  In this paper, we study the mixing angle $\tan \theta_{O\psi}$ and the
mixing amplitude $f_{O\psi}$ between $c \bar c$ and vector glueball $ggg$
in the potential models of heavy quarks and constituent gluons, including 
perturbative dynamics of QCD. 
  From this model calculation, we get the mixing angle at 
$ |\tan \theta_{O\psi}| \approx 0.015$.

  If we take the $J/\psi$ mass at $m_{c \bar c} \approx m_{J/\psi}
\approx 3096 \mbox{ MeV}$ and glueball mass at $m_{ggg} \approx m_{O}
\approx 3168 \mbox{ MeV}$, the mixing amplitude $f_{O\psi}(m_{J/\psi}^2)$, as 
converted from the mixing angle, is equal to $0.008 \mbox{ GeV}^2$.
  In comparison with the phenomenological analysis \cite{Hou:1997qk}, our 
results are off by a factor of two, which lies in the ballpark within our 
approximations.

  It is unlikely to be a fortuitous coincidence that a naive picture of
constituent gluons can give reasonable estimates for both glueball mass  
spectrum and mixing with quarkonium state, as the later quantity is
more sensitive to the actual shape of the glueball wave function.
  At this stage our result seems to be encouraging, and indicates that a
nonrelativistic approximation and more importantly, the constituent gluon   
picture, does capture some grains of truth behind this phenomenological 
puzzle.

\end{document}